\documentclass[twocolumn,aps,prb,showpacs,amsfonts,amssymb,amsmath]{revtex4}

\usepackage{graphicx} 
\usepackage{dcolumn} 
\usepackage{bm} 

\begin{document}
\draft
\title{Nature of Spin Hall Effect in a finite Ballistic
Two-Dimensional System with Rashba and Dresselhaus spin-orbit
interaction}
\author{Yanxia Xing$^1$, Qing-feng Sun$^{1,\ast}$,
and Jian Wang$^{2}$ }
\address{
$^1$Beijng National Laboratory for Condensed Matter Physics and
Institute of Physics, Chinese Academy of Sciences, Bejing 100080,
China\\
$^2$
Department of Physics and the Center of Theoretical and Computational Physics, The University of Hong Kong, Pokfulam Road, Hong Kong, China 
}


\begin{abstract}
The spin Hall effect in a finite ballistic two-dimensional system
with Rashba and Dresselhaus spin-orbit interaction is studied
numerically. We find that the spin Hall conductance is very
sensitive to the transverse measuring location, the shape and size
of the device, and the strength of the spin-orbit interaction. Not
only the amplitude of spin Hall conductance but also its sign can
change. This non-universal behavior of the spin Hall effect is
essentially different from that of the charge Hall effect, in
which the Hall voltage is almost invariant with the transverse
measuring site and is a monotonic function of the strength of the
magnetic field. These surprise behavior of the spin Hall
conductance are attributed to the fact that the eigenstates of the
spin Hall system is extended in the transverse direction and do
not form the edge states.
\end{abstract}


\maketitle

{\sl Introduction:} The Hall effect (from now on we refer as the
charge Hall effect) is a well-known important phenomena in
condensed matter physics. It occurs due to the Lorentz force that
deflects like-charge carriers towards one edge of the sample
creating a voltage transverse to the direction of current.
Recently, another interesting phenomena, the spin Hall effect
(SHE), has been discovered and has attracted considerable
attentions. Here, a spin accumulations emerge on the transverse
sides of the sample, when adding a longitudinal electric field or
bias. If external leads are connected to the sides, the pure
transverse spin current is generated. The SHE can either be
extrinsic due to the spin dependent
scattering\cite{Hirsch,extrinsic} or intrinsic due to the
spin-orbit (SO) interaction. The intrinsic SHE is predicted first
by Murakami {\it et.al.} and Sinova {\sl et.al.} in a Luttinger SO
coupled 3D p-doped semiconductor\cite{Zhangsc} and a Rashba SO
coupled two-dimensional electron gas (2DEG)\cite{Niuq},
respectively. After that a number of recent works have focused on
this interesting issue.\cite{Inoue,Mishchenko,Rashba1,
Yao,Jiang,Chalaev,Bernevig,Sheng,Marinescu,add1,Hankiewicz,
Branislav,Reynoso,Branislav1,add2} For an infinite system, it was
pointed out that the spin Hall conductivity is very sensitive to
disorder,\cite{Inoue,Mishchenko,Rashba1,Yao,Jiang,Chalaev,Bernevig}
and the SHE vanishes even in a very weak disorder. On the other
hand, in the finite mesoscopic ballistic system, the SHE can
survive.\cite{Sheng,Marinescu,add1,Hankiewicz,
Branislav,Reynoso,Branislav1} By using the Landauer-B\"{u}ttiker
formalism and the tight-binding
Hamiltonian\cite{Dattabook,Pareek}, the SHE and spin polarization
have been studied in the dirty\cite{Sheng,Marinescu,add1} or
clean\cite{Hankiewicz,Branislav,Reynoso} mesoscopic samples. These
investigations show that the SHE is still present below a critical
disorder. Experimentally, the SHE is observed on n-type GaAs
\cite{Kato} and on p-type GaAs \cite{Sinova}, where the transverse
spin accumulations are detected by Kerr rotation spectroscopy or
the circularly polarized light-emitting diode, respectively.

In this paper, we study the nature of the SHE in a finite 2DEG,
and mainly focus on the comparison between the SHE and the charge
Hall effect. This is because the SHE and the charge Hall effect
are so analogous, intuitively they should have similar
properties. In the charge Hall effect, the Hall
voltage is a universal constant along the transverse edge, i.e. it is
independent of the transverse measuring location and the width of
device. At least its sign is unchanged. Does the spin current
possess similar universal behavior in SHE?
The results are very surprising and show that the transverse
spin current in the SHE is strongly dependent on the measuring
location and the device's shape.\cite{add3}
Not only the intensity but also
its sign can change. These results indicate that the SHE is not as
clean as the charge Hall effect, and all the measured quantity in
SHE are very sensitive on the details of the system. We attribute
these no-universal behaviors to the extensive eigenstates in the
transverse direction in the SHE.

\begin{figure}
\includegraphics[width=9cm,totalheight=7cm]{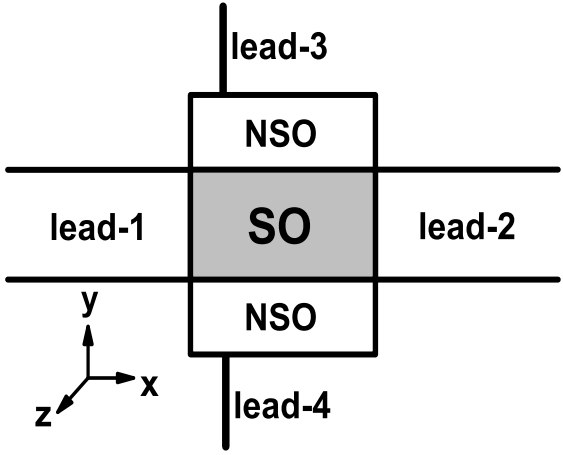}
\caption{ Schematic diagram for the mesoscopic four-terminal
device, which the central gray region (marked by `SO') has the
Rashba and Dresselhaus SO interaction, but two central white zones
(marked by `NSO') are without the SO coupling. }
\end{figure}

{\sl model and  formulation}: The system we considered is shown in
Fig.1 which consists of a finite central ballistic region attached
to four semi-infinite leads. The Rashba and Dresselhaus SO
interactions\cite{Rashba} are present only in the central gray
rectangular region with the size $N\times W$. In order to study
the geometric effect, two zero-SO coupling (NSO) zones (central
white regions in Fig.1) with the size $N\times m$ are also
patched. All the leads are assumed to be clean and ideal, without
any SO coupling. The two longitudinal leads (lead-1 and lead-2)
have the width $W$, which are the same as the width of the central
SO region. On the other hand, in order to study the local spin
Hall conductance and its dependence on the measuring location, two
transverse leads (lead-3 and lead-4) are assumed to be
one-dimensional (1D) with the width $1$, and they can be coupled
to any edge location $S$ along the $x-$direction.

The above system can be described by the Hamiltonian
$H_0=p^2/2m^*+ V(x,y) + \alpha(\sigma_x p_y-\sigma_y
p_x)+\beta(\sigma_x p_x-\sigma_y p_y)$, where $\alpha$ and $\beta$
are the coefficient of the Rashba and Dresselhaus SO
interactions.\cite{Rashba} Then in the tight-binding
representation, this Hamiltonian can be written
as:\cite{Sheng,Marinescu}
\begin{eqnarray}
H&=& \sum\limits_i[ a_{i \uparrow}^\dagger  a_{i
\downarrow}^\dagger ] \left[
\begin{array}{cc}
 -t & -iV_D+V_R \\
       -iV_D-V_R & -t
\end{array}
\right]
\left[
\begin{array}{l}
 a_{i+\delta_x \uparrow} \\ a_{i+\delta_x \downarrow}
\end{array}\right] \nonumber \\
&+&  \sum\limits_i [ a_{i \uparrow}^\dagger  a_{i
\downarrow}^\dagger ] \left[
\begin{array}{cc}
 -t & -iV_R+V_D \\
       -iV_R-V_D & -t
\end{array}
\right]
\left[
\begin{array}{l}
 a_{i+\delta_y \uparrow} \\ a_{i+\delta_y \downarrow}
\end{array}
\right] \nonumber \\
&+&H.C
\end{eqnarray}
where $t=\hbar^2/2m^*a^2$ is the hopping matrix element with the
lattice constant $a$. In order for the band-width of the 1D lead-3
and lead-4 to be in the same range of $-4t$ to $4t$, the hopping
matrix element in these two leads are set to be $2t$. Here
$V_R=\hbar \alpha/2a$ and $V_D=\hbar \beta/2a$ represent the
strength of the Rashba and Dresselhaus interactions, respectively,
and $V_R$ and $V_D$ are non-zero only in the central gray region.
$\delta_x$ and $\delta_y$ in Eq.(1) are the unit vectors along the
$x$ and $y$ directions.

Since there is no SO interactions in the leads, the spin $\sigma$
in the leads are a good quantum number and the definition of the
spin current is unambiguous. Then the particle current
$I_{p\sigma}$ in the lead-$p$ ($p=1$, $2$, $3$, and $4$) with spin
index $\sigma$ ($\sigma=\uparrow$, or $\downarrow$ stands for the $+z$ or $-z$ direction) can be
obtained from the Landauer-B\"{u}ttiker formula: $I_{p\sigma}=
(e/h)\sum_{q\neq p,\sigma'}T_{p\sigma,q\sigma'}(V_p-V_q)$,
\cite{Pareek,Dattabook} where $V_p$ is the bias in the lead-$p$
and $T_{p\sigma,q\sigma'}$ is the transmission coefficient from
the lead-$q$ with spin $\sigma'$ to the lead-$p$ with spin
$\sigma$. The transmission coefficient can be calculated from
$T_{p\sigma,q\sigma'}=Tr[\Gamma_{p\sigma}G^r\Gamma_{q\sigma'}G^a]$,
where the line-width function
$\Gamma_{p\sigma}=i(\Sigma_{p\sigma}^r-\Sigma_{p\sigma}^{r\dagger})$,
the Green's function
$G^r=[G^a]^{\dagger}= \{E_F-H_0-\sum_{p\sigma}\Sigma^r_{p\sigma}\}^{-1}$,
\cite{Dattabook} and $\Sigma_{p\sigma}^r$ is the retarded
self-energy. After solving $I_{p\sigma}$, the spin current $I^s_p$
and the charge current $I^e_p$ can be obtained straightforwardly:
$I^{e}_p = e\{I_{p\uparrow}+I_{p\downarrow}\}$ and $I^{s}_{p} =
(\hbar/2)\{I_{p\uparrow}-I_{p\downarrow}\}$. The terminal voltages
$V_p$ are set as: $V_1= V$ and $V_2=0$,
i.e. a longitudinal bias $V$ is added between the lead-1 and the
lead-2. The transverse lead-3 and lead-4 act as the voltage
probes, and their voltages $V_3$ and $V_4$ are calculated from the
condition $I_3^e=I_4^e=0$. Then the transverse spin Hall
conductances are: $G_{3sH}=I^{s}_3/V_1$ and $G_{4sH}=I^{s}_4/V_1$.
For comparison, we also calculate the transverse charge currents
or the charge conductances ($G_{3e}=I^e_3/V_1$ and
$G_{4e}=I^e_4/V_1$) in the same device but under a different
condition $V_3=V_4=0$ instead of $I_3=I_4=0$. In the numerical
calculation, we take $E_F=-3.8 t$ which is near the band bottom
$-4t$, and $t=1$ as a energy unit, then the corresponding lattice
constant $a \approx 3nm$.\cite{Branislav1} The device's sizes
(i.e. $N$, $W$, and $m$) are chosen in the same order with the
spin precession length $L_{SO}$ over the precessing angle $\pi$.
Here $L_{SO} =\pi a t/2V_R$. If taking $V_R =0.03t$, then
$L_{SO}\approx 50 a$.

\begin{figure}
\includegraphics[width=9.5cm,totalheight=7cm]{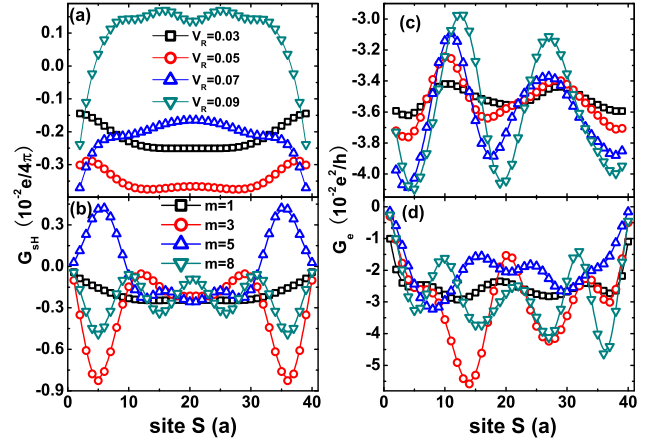}
\caption{ (color online) $G_{sH}$ and $G_e$ {\sl vs} the measuring
site $S$ for different $V_R=0.03$, $0.05$, $0.07$, and $0.09$ with
$m=0$ [panel (a) and (c)], or different width $m$ of the NSO's
zone: $m=1$, $3$, $5$, and $8$ with $V_R=0.03$ [panel (b) and
(d)]. The other parameters are $V_D=0$, $N=40$, and $W=30$. }
\end{figure}

{\sl Numerical results and discussion:} First, we consider the
case that the center region has only Rashba interaction $V_R$
($V_D=0$) and two NSO zones do not exist with $m=0$. While
$V_D=0$, it can be shown that $G_{3sH}=-G_{4sH}\equiv -G_{sH}$ and
$G_{3e}=G_{4e}\equiv G_{e}$. The spin Hall conductance $G_{sH}$
versus the measuring sites $S$ and $V_R$ are depicted in Fig.2a
and Fig.3a. We see that $G_{sH}$ depends on the location of
measuring sites $S$, and it can even change its sign, e.g., when
$V_R=0.09$ (see Fig.2a). On the other hand, at the fixed measuring
site with different $V_R$, the curve of $G_{sH}$ versus $V_R$ can
also cover the range from negative to positive (see Fig.3a). In
contrast to the charge Hall effect, their behaviors are
essentially different. The Hall voltage or the charge Hall
conductance usually are monotonously increasing functions of the
strength of the external magnetic field. Furthermore they usually
are unchanged with the transverse measuring sites.

\begin{figure}
\includegraphics[width=7.5cm,totalheight=7cm]{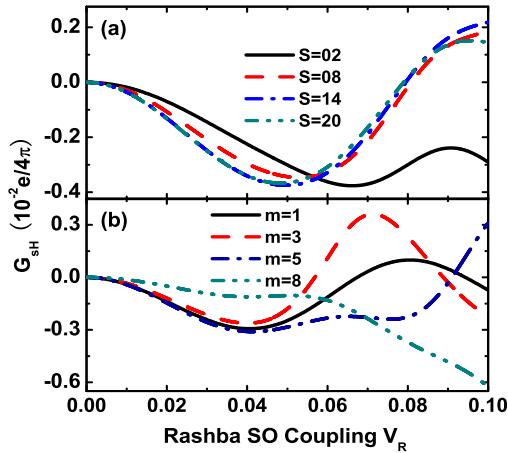}
\caption{ (color online) $G_{sH}$ {\sl vs} $V_R$ for different
site $S$ with $m=0$ [panel (a)], or different width $m$ with
$S=20$ [panel (b)]. The other parameters are same with Fig.2. }
\end{figure}

Next, we attach two NSO zones to the system (see Fig.1). For the
charge Hall effect, the charge accumulates in the transverse
boundaries. If two zones having no magnetic field are attached,
the charge accumulation will naturally transfer from the original
boundaries to the new one, as a result the Hall voltage and the
charge Hall conductance do not change much. How is the spin Hall
conductance $G_{sH}$ affected when two NSO zones are attached?
Fig.2b and Fig.3b show, respectively, $G_{sH}$ versus the site $S$
and $V_R$ for different thickness $m$ of the NSO zone. The results
show that the spin Hall conductance $G_{sH}$ is strongly affected
by the NSO zones. For example, in the curves $G_{sH}$-$S$, for
$m=0$ or $1$ (see Fig.2) $G_{sH}$ is flat, and it is negative at
all site $S$. With increasing $m$, $G_{sH}$ shows an oscillation
behavior. In particular, $G_{sH}$ can be positive, i.e. change its
sign, for some value of $m$ (e.g. $m=5$). In the curve of $G_{sH}$
versus $V_R$ it also exhibits the similar results that $G_{sH}$ is
strongly dependent on $m$ including changing its sign (see
Fig.3b).

For comparison, we also show the charge conductance $G_{e}$ for
the same system but different bias conditions $V_2=V_3=V_4=0$ (see
Fig.2c,d). We see that $G_{e}$ is always negative and exhibits an
oscillation behavior. For $m=0$ (i.e. without the NSO zones),
$G_{e}$ is weakly dependent on the site $S$, whereas for $m\not=
0$, the oscillatory amplitude of $G_{e}$ increases slightly. In
particular, the charge conductance $G_{e}$ is nearly ten times
larger than the spin Hall conductance $G_{sH}$. This also means
$G_{sH}$ is much smaller than the universal value $N e/4\pi$,
where $N$ is the channel number, and $N=2$ in the present device
because that the lead-3's and lead-4's width is $1$. Notice that
in the charge Hall effect, the Hall conductance usually takes the
universal value $N e^2/h$.

\begin{figure}
\includegraphics[width=9.5cm,totalheight=7cm]{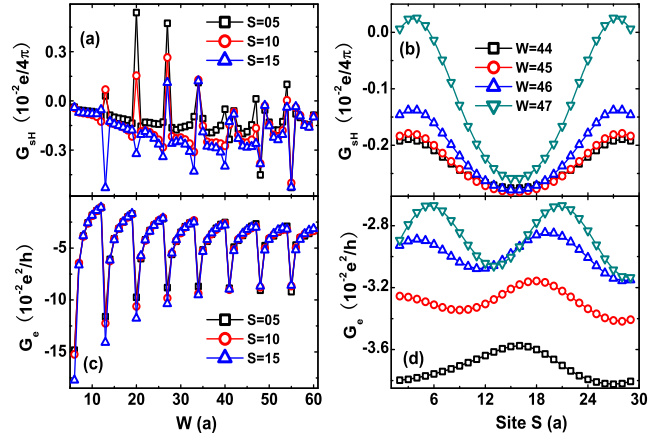}
\caption{ (color online) Left panel:  $G_{sH}$ (a) and $G_e$ (c)
{\sl vs} the transverse width $W$ for the site $S=5$, $10$, and
$15$. Right panel: $G_{sH}$ (b) and $G_e$ (d) {\sl vs} the site
$S$ for different transverse width $W=44$, $45$, $46$, and $47$.
The other parameters are $V_R=0.03$, $V_D=0$, $N=30$, and $m=0$.}
\end{figure}

Let us study the spin Hall conductance $G_{sH}$ versus the
transverse width $W$ of the center SO's regions. $G_{sH}$ and
$G_{e}$ versus $W$ exhibit almost periodic peaks (see Fig.4a,c).
Note that the cutoff energy of the subband (i.e. the transverse
energy levels) are about $n^2\hbar^2 \pi^2/2mW^2$, which shifts
down with increasing the width $W$. For the Fermi level $E_F$
across a subband, a jump emerges in the curves of $G_{sH}$-$W$ (or
$G_{e}$-$W$), due to the large density of state near the band
edge. As a result for a given period (e.g. $W$=44, 45, 46, and
47), $G_{sH}$ and $G_{e}$ versus the site $S$ (see Fig.4b,d),
exhibit the oscillation behavior. As the Fermi level across the
subband edge ($W$=47), $G_{sH}$ can change its sign while $G_e$
are always negative.

\begin{figure}
\includegraphics[width=7.5cm,totalheight=7cm]{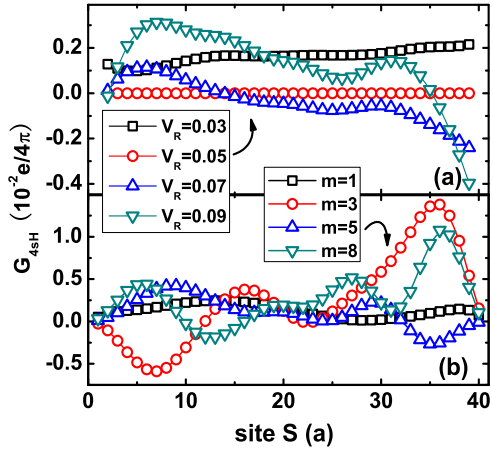}
\caption{ (color online) $G_{4sH}$ {\sl vs} the measuring site $S$
for different Rashba SO coupling strength $V_R=0.03$, $0.05$,
$0.07$, and $0.09$ with $m=0$, or different width of the NSO's
zone: $m=1$, $3$, $5$, and $8$ with $V_R=0.03$. The other
parameters are $V_D=0.05$, $N=40$, and $W=30$.}
\end{figure}

In the following, we investigate the case when the Dresselhaus SO
interaction is present, i.e. $V_D\not=0$. As mentioned above, at
$V_D=0$ the spin currents through the lead-3 and lead-4 are
conserved, i.e., $G_{3sH}= -G_{4sH}$. However, when $V_D \not=0$
and $V_R\not=0$, $G_{3sH} \not= -G_{4sH}$. On the other hand, the
spin Hall conductance has the symmetry with $G_{3sH}(S) = -
G_{4sH}(W-S)$ due to the symmetry of our system.
It is worth to point out when $V_D=V_R$, $G_{4sH}=G_{3sH}=0$, which
is similar with the Ref.(14).
In the Fig.5,
$G_{4sH}$ versus the site $S$ for different $V_R$ or different
width $m$ of the NSO zone are plotted. Here $G_{4sH}$ exhibits
similar characters as in the case of $V_D=0$: $G_{4sH}$ is very
sensitive to the transverse measuring site $S$, and it can even
changes its sign (e.g. $V_R=0.07,0.09$). While $m\not=0$,
$G_{4sH}$ oscillates with the site $S$ along with the variation of
its sign. All these behaviors are in contrast to the charge Hall
conductance.

Finally, we emphasize that the sensitivity of spin Hall
conductance to the location of measuring sites is a generic
feature not due to the 1D nature of the lead-3 and lead-4. We have
performed similar calculations when the width of lead-3 and lead-4
are $3$ and $5$. The conclusion remains. In addition, if the
lead-3 and lead-4 are placed at two different measuring sites along
x-direction, $G_{3sH}$ and $G_{4sH}$ are affected even stronger.

Why are the characters of the SHE so different with the charge
Hall effect? Why is the spin Hall conductance $G_{sH}$ so
sensitive (even its sign) to the measurement site $S$, the
shape of device, and so on? We attribute them to following two
reasons. (1). In the quantum Hall effect the edge states emerge
and play an important role. However for a system that exhibits
SHE, e.g., the quasi 1D quantum wire having Rashba SO interaction,
its eigen states are extended in the transverse direction and they
do not form edge states.\cite{sun} (2). The force in the charge
Hall effect always points to a specific direction, e.g. $+y$. But
the force in the SHE is dependent on the spin $\sigma$, and its
sign can vary.\cite{shen}

In summary, the spin Hall conductance is strongly dependent on the
transverse measuring site, the device's shape, and the strength of
the spin-orbit interaction. Not only the magnitude but also its
sign can change. These characters are very different from that of
the charge Hall effect, and the spin Hall conductance is not
universal as the charge Hall conductance.

{\bf Acknowledgments:} We gratefully acknowledge helpful
discussions with Prof. H. Guo. This work was supported by
NSF-China under Grant Nos. 90303016, 10474125, and 10525418. J.W.
is supported by RGC grant (HKU 7044/05P) from the government SAR
of Hong Kong.

\end{document}